\documentclass[12pt,preprint]{aastex}

\usepackage{amsmath,epsf,multicol,natbib,graphicx}
\usepackage[alpha,z,expert]{tmsfnts}
\usepackage{ulem}
\usepackage{color}
\usepackage{pdfcolmk}
\definecolor{DarkGreen}{rgb}{0.0, 0.5, 0.0}
\definecolor{purple}{rgb}{0.5, 0.0, 0.5}
\definecolor{red}{rgb}{1, 0.0, 0.0}
\definecolor{green}{rgb}{0, 1.0, 0.0}

\newcommand{\sigT}{\mbox{$\sigma_{\mbox{\tiny T}}$}}
\newcommand{\Tcmb}{\mbox{$T_{\mbox{\tiny CMB}}$}}
\newcommand{\kB}{\mbox{$k_{\mbox{\tiny B}}$}}

\newcommand{\Lamee}{\mbox{$\Lambda_{ee}$}}
\newcommand{\rhogas}{\mbox{$\rho_{\mbox{\scriptsize gas}}$}}
\newcommand{\Mgas}{\mbox{$M_{\mbox{\scriptsize gas}}$}}
\newcommand{\Mtot}{\mbox{$M_{\mbox{\scriptsize tot}}$}}
\newcommand{\Yint}{\mbox{$Y_{\mbox{\scriptsize int}}$}}
\newcommand{\Ycyl}{\mbox{$Y_{\mbox{\scriptsize cyl}}$}}
\newcommand{\Ysph}{\mbox{$Y_{\mbox{\scriptsize sph}}$}}
\newcommand{\fgas}{\mbox{$f_{\mbox{\scriptsize gas}}$}}
\newcommand{\LCDM}{\mbox{$\Lambda$CDM}}

\begin{document}
\normalem

\title{\bf Application of a Self-Similar Pressure Profile to 
Sunyaev-Zel'dovich Effect Data from Galaxy Clusters} 

\author{
Tony~Mroczkowski,\altaffilmark{1,2,3}
Max~Bonamente,\altaffilmark{4,5}
John~E.~Carlstrom,\altaffilmark{6,7,8,9}
Thomas~L.~Culverhouse,\altaffilmark{6,7}
Christopher~Greer,\altaffilmark{6,7} 
David~Hawkins,\altaffilmark{10}
Ryan~Hennessy,\altaffilmark{6,7} 
Marshall~Joy,\altaffilmark{5}
James~W.~Lamb,\altaffilmark{10}
Erik~M.~Leitch,\altaffilmark{6,7} 
Michael~Loh,\altaffilmark{6,9}
Ben~Maughan,\altaffilmark{11,12}
Daniel~P.~Marrone,\altaffilmark{6,8,13}
Amber~Miller,\altaffilmark{1,14,15}
Stephen~Muchovej,\altaffilmark{2}
Daisuke~Nagai,\altaffilmark{16,17}
Clem~Pryke,\altaffilmark{6,7,8}
Matthew~Sharp,\altaffilmark{6,9} 
and David~Woody\altaffilmark{10}}

\altaffiltext{1}{Columbia Astrophysics Laboratory, Columbia University, New York, NY 10027}
\altaffiltext{2}{Department of Astronomy, Columbia University, New York, NY 10027}
\altaffiltext{3}{Department of Physics and Astronomy, University of Pennsylvania, Philadelphia, PA 19104}
\altaffiltext{4}{Department of Physics, University of Alabama, Huntsville, AL 35899}
\altaffiltext{5}{Department of Space Science, VP62, NASA Marshall Space Flight Center, Huntsville, AL 35812}
\altaffiltext{6}{Kavli Institute for Cosmological Physics, University of Chicago, Chicago, IL 60637}
\altaffiltext{7}{Department of Astronomy and Astrophysics, University of Chicago, Chicago, IL 60637}
\altaffiltext{8}{Enrico Fermi Institute, University of Chicago, Chicago, IL 60637}
\altaffiltext{9}{Department of Physics, University of Chicago, Chicago, IL 60637}
\altaffiltext{10}{Owens Valley Radio Observatory, California Institute of Technology, Big Pine, CA 93513} 
\altaffiltext{11}{Department of Physics, University of Bristol, Tyndall Ave, Bristol BS8 1TL, UK.}
\altaffiltext{12}{Harvard-Smithsonian Center for Astrophysics, 60 Garden St., Cambridge, MA 02138}
\altaffiltext{13}{Jansky Postdoctoral Fellow, National Radio Astronomy Observatory}
\altaffiltext{14}{Department of Physics, Columbia University, New York, NY 10027}
\altaffiltext{15}{Alfred P. Sloan Fellow}
\altaffiltext{16}{Department of Physics, Yale University, New Haven, CT 06520}
\altaffiltext{17}{Yale Center for Astronomy \& Astrophysics, Yale University, New Haven, CT 06520}

\begin{abstract} 
 
We investigate the utility of a new, self-similar pressure profile for
fitting Sunyaev-Zel'dovich (SZ) effect observations of galaxy
clusters.  Current SZ imaging instruments---such as the Sunyaev-Zel'dovich Array
(SZA)---are capable of probing clusters over a large
range in physical scale.  A model is therefore required that can accurately describe a
cluster's pressure profile over a broad range of radii, from 
the core of the cluster out to a significant fraction of the virial radius.  
In the analysis presented here, we fit a
radial pressure profile derived from simulations and
detailed X-ray analysis of relaxed clusters to SZA observations of
three clusters with exceptionally high quality X-ray data: A1835,  A1914, and CL~J1226.9+3332.  From the
joint analysis of the SZ and X-ray data, we derive physical
properties such as gas mass, total mass, gas fraction and the intrinsic, integrated
Compton $y$-parameter. 
We find that parameters derived from the joint fit to the SZ and X-ray data
agree well with a detailed, independent X-ray-only analysis of the
same clusters. 
In particular, we find that, when combined with X-ray imaging data,
this new pressure profile yields an independent electron radial temperature
profile that is in good agreement with spectroscopic X-ray
measurements.

\end{abstract}



\section{Introduction}

The expansion history of the universe and the growth of large-scale
structure are two of the most compelling topics in cosmology.  As
galaxy clusters are the largest collapsed objects in the universe,
taking a Hubble time to form, their abundance and evolution are
critically sensitive to the details of that expansion history.
Cluster surveys can therefore provide fundamental clues to the nature
and abundance of dark matter and dark energy \citep[see,
e.g.,][]{white1993b,frenk1999,haiman2001,weller2002}.

While clusters have been extensively studied using X-ray observations
of the hot gas that comprises the intracluster medium (ICM), radio
measurements of the same gas via the Sunyaev-Zel'dovich (SZ) effect \citep{sunyaev1972}
provide an independent and complementary probe of the ICM \cite[e.g.,][]{carlstrom2002}.  The SZ effect
arises from inverse Compton scattering of cosmic microwave background
(CMB) photons by the electrons in the ICM, imparting a detectable
spectral signature to the CMB that is independent of the redshift of
the cluster. As the SZ effect is a measure of the integrated line-of-sight
electron density, weighted by temperature, i.e., the integrated
pressure (see \S~\ref{xray_sze}), it probes different properties than
the X-ray emission, which is proportional to the square of the
electron density. Cluster surveys exploiting the redshift independence
of the SZ effect are now being conducted by a variety of instruments,
including the SZA \citep{muchovej2007}, the South Pole Telescope
\citep{ruhl2004}, the Atacama Cosmology Telescope \citep{kosowsky2003}
and the Atacama Pathfinder Experiment (APEX) SZ instrument \citep{dobbs2006}.

To maximize the utility of clusters as cosmological probes we must
understand how to accurately relate their observable properties to
their total masses.  The integrated SZ signal from a cluster is
proportional to the total thermal energy of the cluster, and is
therefore a measure of the underlying gravitational potential, and
ultimately the dark matter content, within a cluster.
The SZ flux of a cluster thereby should provide a robust, low scatter
proxy for the total cluster mass, \Mtot\
\citep[see, for example,][]{dasilva2004,motl2005,nagai2006,reid2006}.

To date, cosmological studies combining SZA and X-ray data have relied
almost exclusively on the isothermal $\beta$-model
\citep[first used in][]{cavaliere1976,cavaliere1978} to fit the SZ signal in the
region interior to $r_{2500}$
\citep[see][for applications of the isothermal $\beta$-model to SZ+X-ray data]{grego2000,
reese2002,laroque2006,bonamente2006,bonamente2008}.
Here $r_{2500}$ is the radius within which the mean cluster density is a
factor of 2500 over the critical density of the universe at the
cluster redshift.  While the isothermal $\beta$-model recovers global
properties of clusters quite accurately in this regime
\citep{laroque2006}, deep X-ray observations of nearby
clusters show that isothermality is a poor description of the cluster
outskirts ($r\sim r_{500}$) \citep[see e.g.][and references therein]{piffaretti2005,vikhlinin2005a,pratt2007}.  An improved model
for the cluster gas, accurate to large radii, is therefore critical for
the analysis and cosmological interpretation of SZ data obtained with
new instruments that are capable of probing the outer regions ($r \sim r_{500}$) of clusters.
Such a model must be simple enough that it can be constrained by SZ
data with limited angular resolution and sensitivity typical of data sets acquired by SZ survey
instruments optimized for detection, rather than imaging.  While the
$\beta$-model has the virtue of simplicity, previous attempts to relax the
assumption of isothermality typically required high-significance,
spatially-resolved X-ray spectroscopy; such data are seldom obtained in
short X-ray exposures of high-redshift clusters. This is particularly 
true for the cluster outskirts \citep[see, e.g.,][]{laroque2006}.  
Attempts to move beyond the $\beta$-model have typically improved the 
modeling of the gas density only within the core ($r\lesssim0.15\,r_{500}$).

In this work, we investigate a new model for the cluster gas pressure 
by using it to fit SZ data from the SZA and X-ray data from {\em Chandra}
to three well-studied clusters: Abell 1835,  Abell 1914, and CL~J1226.9+3332. The model was
derived by \citet{nagai2007b} from simulations and from detailed
analysis of deep {\it Chandra} measurements of nearby relaxed
clusters.  The simplicity of this model---and the fact that SZ data
are inherently sensitive to the integrated electron pressure---allow
it to be used either in conjunction with X-ray imaging data, or fit to
SZ data alone.  The outline of the paper is as follows: in
\S~\ref{modeling}, we present the details of the model, and couple it
with an ICM density model that allows the inclusion of X-ray imaging
data.  In \S~\ref{realobs}, we apply this method to three clusters,
combining new data from the SZA with {\em Chandra} X-ray imaging
data.  We demonstrate the utility of this model by applying it to
SZ+X-ray data without relying on X-ray spectroscopic information.
Results from the joint SZ+X-ray analysis are then compared to results
from an X-ray-only analysis, including spectroscopic data, in \S~\ref{results}.
We offer our conclusions in \S~\ref{conclusions}.


\section{Cluster Gas Models}\label{modeling}

\subsection{Sunyaev Zel'dovich Effect and X-ray Emission}\label{xray_sze}

The thermal SZ effect is a small ($< 10^{-3}$) distortion 
in CMB intensity caused by inverse Compton scattering of CMB 
photons by energetic electrons in the hot intracluster gas
\citep{sunyaev1970,sunyaev1972}.  This spectral distortion can be
expressed, for dimensionless frequency $x \equiv h\nu/\kB \Tcmb$, where
 $h$ is Planck's constant, $\nu$ is frequency, $k_B$ is Boltzmann's constant, and 
\Tcmb\ is the primary CMB temperature, as
 the change $\Delta I_{SZ}$ relative to the primary CMB intensity normalization $I_0$,
\begin{eqnarray}
\label{eq:thermal_sz}
\frac{\Delta I_{SZ}}{I_0} &=& \frac{k_B \, \sigma_T}{m_e c^2} \int \!\!  g(x,T_e) \, n_e T_e \,d\ell \\
\label{eq:thermal_sz2}
&=& \frac{\sigma_T}{m_e c^2} \int \!\! g(x,T_e) \, P_e \,d\ell.
\end{eqnarray}
Here \sigT\ is the Thomson scattering cross-section of the electron,
$\ell$ is the line of sight, and $m_e c^2$ is an electron's rest energy.
The factor $g(x,T_e)$ encapsulates the frequency dependence of the SZ effect intensity.
For non-relativistic electrons, 
\begin{equation}
\label{eq:g_x}
g(x) = \frac{x^4 e^x}{(e^x-1)^2} \left(x \frac{e^x + 1}{e^x - 1} - 4\right).
\end{equation}
We use the \citet{itoh1998} relativistic corrections to Eq.~\ref{eq:g_x}, which are appropriate
for fitting thermal SZ observations, and discuss their impact on the fit ICM profiles in \S~\ref{profiles}.
Note that we have used the ideal gas law ($P_e = n_e k_B T_e$)
to obtain Eq.~\ref{eq:thermal_sz2} from Eq.~\ref{eq:thermal_sz}; we use
this to relate SZ intensity directly to the ICM electron pressure. 

The X-ray emission from massive clusters arises 
predominantly as thermal bremsstrahlung from electrons.  The
X-ray surface brightness produced by a cluster at redshift $z$ is given by \citep[e.g.,][]{sarazin1988}
\begin{equation}
\label{eq:xray_sb}
S_X = \frac{1}{4\pi (1+z)^4} \! \int \!\! n_e^2 \Lamee(T_e,Z) \,d\ell,
\end{equation}
where the integral is evaluated along the line of sight and the X-ray
cooling function $\Lamee$ is proportional to $T_e^{1/2}$.  The X-ray
emission is therefore proportional to the square of the gas density,
with a relatively weak dependence on $T_e$.  Separate spectroscopic
observations of X-ray line emission can be used to measure the gas
temperature.  Throughout this work, we use the Raymond-Smith plasma
emissivity code \citep{raymond1977} with constant metallicity $Z=0.3 Z_{\odot}$,
yielding $\mu_e=1.17$ and $\mu = 0.61$ for the mean molecular weight
of the electrons and gas, respectively, using the elemental abundances provided by \citet{anders1989}.  
The SZ and X-ray observables $\Delta I_{SZ}$ and $S_X$ are computed by evaluating the integrals
in Eqs.~\ref{eq:thermal_sz2} \& \ref{eq:xray_sb} numerically.

\subsection{Three-dimensional Models of the ICM Profiles}
\label{models}

In this work, we adopt an analytic parameterization of the cluster radial
pressure profile proposed by \citet{nagai2007b}~(hereafter N07),
\begin{equation}
P_e(r) = \frac{P_{e,i}}{(r/r_p)^c 
\left[1+(r/r_p)^a\right]^{(b-c)/a}},
\label{eq:press}
\end{equation}
where $P_{e,i}$ is a scalar normalization of the pressure profile,
$r_p$ is a scale radius (typically $r_p \approx r_{500}/1.3$), and the
parameters $(a,b,c)$ respectively describe the slopes at intermediate
($r\approx r_p$), outer ($r>r_p$), and inner ($r \ll r_p$) radii.
Note that Eq.~\ref{eq:press} is a generalization of the analytic
fitting formula obtained in numerical simulations as a parameterization of
the distribution of mass in a dark matter halo \citep[][NFW profile]{navarro1997}.  
This choice is reasonable because
the gas pressure distribution is primarily determined by the dark
matter potential.  The use of a parameterized
pressure profile is further motivated by the fact that self-similarity
is best preserved for pressure, as demonstrated by the low
cluster-to-cluster scatter seen when using these parameters to fit
numerical simulations.  The NFW profile -- in its pure form -- has been applied by 
\citet{atriobarandela2008} to fit the electron pressure profiles of SZ observations of clusters, 
who demonstrated an improvement over the application of the isothermal $\beta$-model to SZ cluster studies.
In this work, we adopt the fixed slopes
$(a,b,c)=(0.9,5.0,0.4)$, which N07 found to closely match the observed profiles of
the \emph{Chandra} X-ray clusters and the results of numerical simulations
in the outskirts of a broad range of relaxed clusters.\footnote{The original values
published in N07 were $(a,b,c)=(1.3,4.3,0.7)$.  These have recently been updated, however,
and will be published in an erratum to N07.}

The density model used to fit the X-ray image data is a
simplified version of the model employed by \cite{vikhlinin2006}~(hereafter V06) is
\begin{equation}
    n_e^2(r) = n_{e0}^2\;\frac{(r/r_c)^{-\alpha}}{\left[1+(r/r_c)^2\right]^{3\beta-\alpha/2}}\\
                 \frac{1}{\left[1+(r/r_s)^\gamma\right]^{\varepsilon/\gamma}}\\
                + \frac{n_{e02}^2}{\left[1+(r/r_{c2})^2\right]^{3\beta_2}}.
\label{eq:Vikhlinin}
\end{equation}
We refer to Eq.~\ref{eq:Vikhlinin}, used in the independent X-ray analysis
to which we compare our SZ+X-ray results, as the ``V06 density model.''  Since the cluster core
contributes negligibly to the SZ signal observed by the SZA, we 
exclude the inner 100~kpc of the cluster from the X-ray images used in
the joint SZ+X-ray analysis.  Recognizing $\alpha$ as the component introduced
by \cite{pratt2002} to fit the inner slope of a cuspy cluster density
profile, and that the second $\beta$-model component ($\beta_2$) is present
explicitly to fit the cluster core, we simplify the V06 density model
to
\begin{equation}
n_e(r) = n_{e0}
\left[1+(r/r_c)^2\right]^{-3\beta/2}
\left[1+(r/r_s)^{\gamma}\right]^{-0.5\varepsilon/\gamma},
\label{eq:svm}
\end{equation}
where $r_c$ is the core radius, and $r_s$ is the radius at which the
density profile steepens with respect to the traditional
$\beta$-model.  Following V06, we fix $\gamma=3$, as this provides an
adequate fit to all clusters in the V06 sample.  We refer to this
model as the ``Simplified Vikhlinin Model'' (SVM).  Note that in the
limit $\varepsilon \rightarrow 0$, Eq.~\ref{eq:svm} reduces to
the standard $\beta$-model.  The SVM is thus a
modification to the $\beta$-model that has the additional freedom to
extend from the core ($r \gtrsim r_c$) to the outer regions of cluster gas, 
spanning intermediate ($r \gtrsim r_s$) to large radii ($r \sim r_{500}$).

With the electron pressure and density in hand, we may also derive the
electron temperature of the ICM using the ideal gas law,
$T_e(r)=P_e(r)/k_B n_e(r)$, where $P_e$ and $n_e$ are given by
Eqs.~\ref{eq:press} and \ref{eq:svm}, respectively.  Note that $T_e(r)$
derived in this way is used in the analysis of X-ray surface brightness
(Eq.~\ref{eq:xray_sb}).  
Hereafter, we refer to this 
jointly-fit cluster gas model as the N07+SVM profile.

For comparison with previous work \citep[e.g.][]{bonamente2008}, we
also employ the isothermal $\beta$-model for joint analysis of the SZ
and X-ray data. In this model the density is given by Eq.~\ref{eq:svm}
with $\varepsilon=0$ and $T_e(r)$ is a constant equal to the
spectroscopically-measured temperature, $T_X$. The shape parameters of
the isothermal $\beta$-model, $r_c$ and $\beta$, are jointly fit to
the SZ and X-ray data, while the X-ray surface brightness
(Eq.~\ref{eq:xray_sb}) and SZ intensity profile
(Eq.~\ref{eq:thermal_sz2}) normalizations are independently determined
from the X-ray and SZ data, respectively.

\subsection{Parameter Estimation Using the Markov Chain Monte Carlo Method \label{mcmc}}
Our models have five free parameters to describe the radial distribution of the gas density (see
Eq.~\ref{eq:svm}) and two parameters for the electron
pressure (see Eq.~\ref{eq:press}).  Additional parameters such as the
cluster centroid, X-ray background level, as well as the positions, fluxes
and spectral indices of compact radio sources are also included
where necessary.  The Markov chain Monte Carlo (MCMC) method is used to
extract the model parameters from the SZ and X-ray data, as described by \citet{bonamente2004}. 
In this section we provide a brief overview of
this method, focusing on the changes to accommodate the N07 pressure model.

The first step in fitting the SZ data is to compute the model image over
a regular grid, sampled at less than half the
smallest scale the SZA can probe.  This image is multiplied by the 
primary beam of the SZA, transformed via FFT to Fourier space (where the data 
are naturally sampled by an interferometer; see \S~\ref{sze_visfits}), and interpolated to the Fourier-space
coordinates of the SZ data.  The likelihood function for the SZ data is
then computed directly in the Fourier plane, where the noise properties of 
the interferometric data are well-characterized.

The first step of the MCMC method is the calculation of the joint
likelihood $\mathcal{L}$ of the X-ray and SZ data with the model.
The SZ likelihood is given by
\begin{equation}
\ln(\mathcal{L}_{SZ})=\displaystyle \sum_i \left[-\frac{1}{2} \left(\Delta R_i^2+\Delta I_i^2\right)\right] W_i,
\end{equation}
where $\Delta R_i$ and $\Delta I_i$ are the differences between model
and data for the real and imaginary components at each point $i$ in
the Fourier plane, and $W_i$ is a measure of the Gaussian noise
($1/\sigma^2$).

Since the X-ray counts, treated in image space, are distributed
according to Poisson statistics, the likelihood of the model fit is given by
\begin{equation}
\ln(\mathcal{L}_{image})=\displaystyle \sum_i \left[ D_i \ln(M_i) - M_i -\ln(D_i!) \right],
\label{eq:L_image}
\end{equation}
where $M_i$ is the model prediction (including cluster and background
components), and $D_i$ is the number of counts detected in pixel
$i$. The inner 100~kpc of the X-ray images---as well as any detected X-ray point sources---are
 excluded from the fits by excluding these regions from the calculation
of $\ln(\mathcal{L}_{image})$ in Eq.~\ref{eq:L_image}.

The joint likelihood of the spatial and spectral models is given by
$\mathcal{L}=\mathcal{L}_{SZ}\mathcal{L}_{Xray}$. For the N07+SVM
fits, $\mathcal{L}_{Xray}$ is simply $\mathcal{L}_{image}$. Following
\citet{bonamente2004,bonamente2006}, the X-ray likelihood for the
$\beta$-model fits is $\mathcal{L}_{image}\mathcal{L}_{Xspec}$, as
these must incorporate the likelihood $\mathcal{L}_{Xspec}$ of the
spectroscopic determination of $T_X$. The likelihood is used to
generate the Markov parameter chains, and convergence of the chain to
a stationary distribution is established using the Raftery-Lewis and Geweke tests
\citep{raftery1992,gilks1996}.

\subsection{Calculation of \Mgas, \Mtot, and \Yint}\label{derived}
With knowledge of the three-dimensional gas profiles, we compute
global properties of galaxy clusters as follows.  The gas mass
$\Mgas(r)$ enclosed within radius $r$ is obtained by integrating the
gas density $\rhogas \equiv \mu_e m_p n_e(r)$ over a spherical
volume:
\begin{equation}
\Mgas(r) = 4 \pi \! \int_{0}^{r} \!  \rhogas (r') r'^2 dr'.
\label{eq:gasmass}
\end{equation}
The total mass, \Mtot, can be obtained by solving the hydrostatic
equilibrium equation as:
\begin{equation}
\Mtot(r) = -\frac{r^2}{G \rhogas(r)} \frac{dP(r)}{dr},
\label{eq:hse}
\end{equation}
where $P=(\mu_e/\mu)P_e$ is the total gas pressure.  We then compute
the gas mass fractions as $\fgas =\Mgas/\Mtot$.  

The line of sight Compton $y$-parameter, which characterizes the strength of 
Compton scattering by electrons, is defined
\begin{equation}
y \equiv \frac{k_B \sigma_T}{m_e c^2} \int \!\! n_e T_e \,d\ell.
\end{equation}
We compute the volume-integrated Compton $y$-parameter, $Y$, from the
pressure profile fit to the SZ observations for both
cylindrical and spherical volumes of integration. The
cylindrically-integrated quantity, \Ycyl, is calculated within an
angle $\theta$ on the sky, corresponding to physical radius $R=\theta d_A$ at the
redshift of the cluster,
\begin{equation}
\Ycyl\left(R\right) \equiv 2\pi\,{d_A^2} \int_0^\theta \!\! y\left(\theta\right) \, \theta' d\theta' =
2\pi \int_0^R \!\! y\left(r'\right) \, r' dr' = \frac{2\pi\,\sigma_T}{m_e \, c^2} 
\int_{-\infty}^{\infty} \!\! d\ell \int_0^R \! P_e\left(r'\right) \, r' dr'. 
\label{eq:Ycyl}
\end{equation}
The last form makes explicit the infinite limits of integration in the
line of sight direction, originating with the definition of $y$
(Eqs.~\ref{eq:thermal_sz} and \ref{eq:thermal_sz2}). The
spherically-integrated quantity, \Ysph, is obtained by integrating the
pressure profile within a radius $r$ from the cluster center,
\begin{equation}
\Ysph(r) = \frac{4 \pi \,\sigma_T}{m_e \, c^2} \int_{0}^{r} \!\! P_e (r') \, r'^2 dr.
\label{eq:Ysph}
\end{equation}
The parameter $\Ysph(r)$ is thus proportional to the thermal energy content of the ICM.

To compute the global cluster properties described above, one needs to
define a radius out to which all quantities will be calculated.
Following \citet{laroque2006} and \citet{bonamente2006}, we compute global
properties of clusters enclosed within the overdensity radius $r_{\Delta}$,
within which the average density $\langle \rho \rangle$ of the cluster is a specified
fraction $\Delta$ of the critical density, via
\begin{equation}
\frac{4}{3} \pi \, \rho_c(z) \, \Delta \, r_{\Delta}^3 = \Mtot(r_{\Delta}),
\label{rvir}
\end{equation}
where $\rho_c(z)$ is the critical density at cluster redshift $z$,
and $\Delta \equiv \langle \rho \rangle /\rho_c(z)$. 
In this work, we evaluate cluster properties at density
contrasts of $\Delta=2500$ and $\Delta=500$. 
The overdensity radius $r_{2500}$ has been used in previous OVRO and BIMA 
interferometric SZ studies \citep[e.g.][]{laroque2006,bonamente2008} as well as in many 
X-ray cluster studies \citep[e.g.][]{vikhlinin2006, allen2007}, while $r_{500}$ is a radius
reachable with SZA and deep \emph{Chandra} X-ray data.


\section{Data\label{realobs}}

\subsection{Cluster Sample \label{clusterselection}}

For this work, we selected three massive clusters that are well studied at
X-ray wavelengths, and span a range of redshifts
($z$~=~0.17--0.89) and morphologies, on which to test the
joint analysis of the {\em Chandra} X-ray and SZA data.  
We assume a \LCDM\ cosmology with $\Omega_M=0.3$, $\Omega_\Lambda=0.7$, and $h=0.7$ 
throughout our analysis.

Located at $z=0.25$, Abell 1835 (A1835) is an intermediate-redshift, relaxed
cluster, as evidenced by its circular morphology in the X-ray images
and its cooling core \citep{peterson2001}.  To demonstrate the
applicability of our technique for high redshift clusters, we analyzed
CL~J1226.9+3332 (CL1226), an apparently relaxed cluster at $z=0.89$
\citep{maughan2004b,maughan2007b}.  To assess how this method performs
on somewhat disturbed clusters, we also analyzed Abell 1914 (A1914), an
intermediate redshift ($z=0.17$) cluster with a hot subclump near the core.
When the subclump is not excluded from the X-ray analysis,
\citet[][hereafter M08]{maughan2008} find a large X-ray centroid shift
in the density profile, which they use as an indicator of an unrelaxed
dynamical state.

In the following sections, we discuss the instruments, data
reduction, and analysis of the SZ and X-ray data. Details of these observations, 
including the X-ray fitting regions, the unflagged, on-source 
integration times, and the pointing centers used for the SZ observations, are 
presented in Tables \ref{obsTable} and \ref{xrayTable}.
We also review an independent, detailed X-ray-only analysis, with which 
we compare the results of our joint SZ+X-ray analyses.

\subsection{Sunyaev-Zel'dovich Array Observations\label{sza}}

The Sunyaev-Zel'dovich Array is an interferometric array comprising eight 3.5-meter telescopes, 
and is located at the Owens Valley Radio Observatory. 
For the observations presented here, the instrument was configured to operate in an 8-GHz-wide band 
covering 27--35~GHz using the 26--36~GHz receivers
(hereafter referred to as the ``30-GHz'' band) or covering 90--98~GHz using the 80--115~GHz receivers
(the ``90-GHz'' band). 
See \citet{muchovej2007} and Marrone et al. (in preparation) respectively  
for more details about commissioning observations performed with the 
30-GHz and the 90-GHz SZA instruments. Details of the observations presented here, including the sensitivity and
effective resolution (the {\it synthesized beam}) of the long and
short baselines, are given in Table \ref{obsTable}.

An interferometer directly measures the amplitude and phase of Fourier
modes of the sky intensity, with sensitivity to a range of angular
scales on the sky given by $\sim\lambda/B$, where $B$ is the projected separation
of any pair of telescopes, i.e., a {\it baseline}. The
field of view of the SZA is given by the primary beam of a
single telescope, approximately 10.7\arcmin\ at the
center of the 30-GHz band. At 30~GHz, optimal detection of
the arcminute-scale bulk SZ signal from clusters requires the short
baselines of a close-packed array configuration; six of the SZA
telescopes were arranged in this configuration for the observations
presented here, yielding 15 baselines
with sensitivity to $\sim$ 1--5\arcmin\ scales. The two outer antennas,
identical to the inner six, provide an additional 13 long baselines in this configuration,
with sensitivity to small-scale structure, allowing simultaneous
measurement of compact radio sources unresolved by the long
baselines (angular size $\lesssim 20\arcsec$) which could otherwise contaminate the SZ
signal. Observations at 90~GHz with the SZA probe scales at three times the resolution of the 
30-GHz observations for the same array configuration. The short baselines of the 
90-GHz observations thereby bridge the gap between long and short baseline coverage at 
30-GHz.

SZA data are processed in a complete pipeline for the reduction and
calibration of interferometric data, developed within the SZA
collaboration.  Absolute flux calibrations are derived from
observations of Mars, scaled to the predictions
of \citet{rudy1987}.  Data from each observation are bandpass-calibrated
using a bright, unresolved, flat-spectrum radio source, observed at
the start or end of an observation.  The data are regularly
phase-calibrated using radio sources near the targeted field; these
calibrators are also used to track small variations in the antenna
gains.  Data are flagged for corruption due to bad weather, sources of
radio interference and other instrumental effects that could impact
data quality.  A more detailed account of the SZA data reduction
pipeline is presented in \citet{muchovej2007}.

In the SZA cluster observations presented here, the A1835 field
contains three detectable compact sources at 31~GHz: a $2.8\pm0.3$~mJy
central source, a $1.1\pm0.4$~mJy source about one arcminute from
the cluster center, and a $0.8\pm0.4$~mJy source 5.5 arcminutes from center.  
The positions of these sources are
in good agreement with those from the NVSS (which only contains the
central source) and the FIRST surveys.   
The SZA observation of CL1226 contains one detectable compact source,
identified in both FIRST and NVSS, with flux at 31~GHz of $3.9\pm0.2$~mJy.
This source is 4.3 arcminutes from the cluster center \citep[see also][]{muchovej2007} 
and therefore lies outside the field of view of the SZA 90~GHz observations.   
Two compact sources, with positions constrained by NVSS and FIRST, were detected 
at 31~GHz in the A1914 field.  The fluxes of these sources are $1.8\pm0.4$~mJy and
$1.2\pm0.3$~mJy. The stronger was detected in both the NVSS and the FIRST surveys,
while the weaker was only detected in the FIRST survey.  
Table \ref{ptsrcTable} summarizes the properties of the compact radio 
sources detected in the SZA observations.

When fitting compact sources detected in the SZ observations, we calculated
the spectral indices from the measured flux at 31~GHz and 1.4~GHz,
where the latter was constrained by either the NVSS \citep{condon1998} or 
FIRST \citep{white1997} survey, respectively.  
The source fluxes and
approximate coordinates are first identified using the interferometric
imaging package Difmap \citep{shepherd1997}.  We first refine the source
positions by fitting a model in the MCMC routine, and then fix these
positions to their best-fit values when fitting the cluster SZ model.
We leave the source flux a free parameter, so that the cluster SZ
flux and any compact sources are fit simultaneously.

\subsection{X-ray Observations \label{xrayobs_details}}
\label{sec:xray}

All X-ray imaging data used in this analysis were obtained with the {\em
Chandra} ACIS-I detector, which provides spatially resolved X-ray
spectroscopy and imaging with an angular resolution of
$\sim0.5\arcsec$ and energy resolution of $\sim$ 100--200~eV. 
Table~\ref{xrayTable} summarizes the X-ray observations of individual clusters.

For the X-ray data used in the joint SZ+X-ray analysis, images
were limited to 0.7--7~keV in order to exclude the
data most strongly affected by
background and by calibration uncertainties.  The X-ray images---which
primarily constrain the ICM density profiles---were
binned in 1.97$\arcsec$ pixels; this binning sets the limiting
angular resolution of our processed X-ray data, as the {\em Chandra} point
response function in the center of the X-ray image is smaller than our
adopted pixel size.  The X-ray background was measured for each
cluster exposure, using source-free, peripheral regions of the adjacent detector
ACIS-I chips.  Additional details of the {\em
Chandra} X-ray data analysis are presented in \citet{bonamente2004,bonamente2006}.

In \S~\ref{profiles} \& \ref{global}  we compare the results of our joint SZ+X-ray 
analysis to the results of independent X-ray analyses.  For A1914 and CL1226
we use the data and analysis described in detail in M08 and
\cite{maughan2007b} (hereafter M07), respectively.  For A1835, the
ACIS-I observations used became public after  the M08 sample was published, and we 
therefore present its analysis here for the first time.  
The observation of A1835 was calibrated and
analyzed using the same methods and routines described in M08, which we now 
briefly review.

In the X-ray-only analyses, blank-sky fields are used to 
estimate the background for both the imaging and spectral analysis. The
imaging analysis (primarily used to obtain the gas emissivity profile)
is performed in the 0.7--2~keV energy band to maximize signal to noise.
Similar to the joint SZ+X-ray analyses, these images were also binned into 1.97$\arcsec$ pixels.

For the spectral analysis, spectra extracted from a region of interest were
fit in the 0.6--9~keV band with an absorbed, redshifted
Astrophysical Plasma Emission Code (APEC) \citep{smith2001}
model. This absorption was fixed at the Galactic value.
This spectral analysis was used to derive both the global temperature 
$T_X$, determined within the annulus $r \in [0.15,1.0] \, r_{500}$, and the radial temperature 
profile fits of the V06 temperature profile, given by
\begin{equation}
  T_{\mathrm{3D}}(r) =  T_0
	\left[\frac{(r/r_{\text{cool}})^{a_{\text{cool}}}+T_{\text{min}}/T_0}
	     {(r/r_{\text{cool}})^{a_{\text{cool}}}+1}\right]
	\left[\frac{(r/r_t)^{-a}}{(1+(r/r_t)^b)^{c/b}}\right].
\label{eq:V06_tprof}
\end{equation}
We refer the reader to V06 for details, but note that this temperature profile
is the combination of a cool core component (first set of square brackets, where
the core temperature declines to $T_{\text{min}}$) and a decline at large radii 
(second set of square brackets, where temperature falls at $r \gtrsim r_t$).

An important consideration when using a blank-sky background method is
that the count rate at soft energies can be significantly different in
the blank-sky fields than in the target field, due to differences between
the level of the soft Galactic foreground emission in the target field
and that in the blank-sky field.
This was accounted for in the imaging analysis by normalizing the
background image to the count rate in the target image in regions
far from the cluster center. In the spectral analysis, this was modeled by an 
additional thermal component that was fit
to a soft residual spectrum (the difference between spectra extracted in
source free regions of the target and background datasets; see \citet{vikhlinin2005a}). 
The exception to this was the \emph{XMM-Newton} data
used in addition to the {\em Chandra} data for CL1226. As discussed
in M07, a local background was found to be more reliable
for the spectral analysis in this case, thus requiring no correction for
the soft Galactic foreground.

The M07/M08 X-ray analysis methods exploit the full V06 density and temperature models 
(Eqs.~\ref{eq:Vikhlinin} \& \ref{eq:V06_tprof}, respectively) to fit the emissivity and temperature profiles derived for each cluster, 
and the results of these fits are used to derive the total hydrostatic mass profiles of each system. 
Uncertainties for the independent, X-ray-only analysis method are derived using a 
Monte Carlo randomization process.  These fits involved typically
$\sim$1000 realizations of the temperature and surface brightness profiles, fit
to data randomized according to the measured noise.
We refer the reader to M07 and M08, where this fitting procedure is described in detail.

\section{Results \label{results}}

\subsection{SZ Cluster Visibility Fits}\label{sze_visfits}

Interferometric SZ data are in the form of visibilities $V_\nu(u,v)$ \citep[see, for example,][]{thompson2001},
which for single, targeted cluster observations with the SZA can be expressed  in the small angle approximation as
\begin{equation}
 V_\nu(u,v) = \int \! \int A_\nu(x,y) \, I_\nu(x,y) \, {e^{- i 2 \pi (ux + vy)}} \, dx \, dy.
\label{eq:visibility}
\end{equation}
Here $u$ and $v$ (in number of wavelengths) are the Fourier conjugates of the direction cosines 
$x$ and $y$ (relative to the observing direction), $A_\nu(x,y)$ is the angular power sensitivity pattern of each antenna at frequency $\nu$,
and $I_\nu(x,y)$ is the intensity pattern of the sky (also at $\nu$).
Eq.~\ref{eq:visibility} is recognizable as a 2-D Fourier transform, so the visibilities
give the flux for the Fourier mode for the corresponding $u,v$-coordinate.

By combining Eqs.~\ref{eq:thermal_sz} and \ref{eq:visibility}, we can remove the frequency
dependence from the measured cluster visibilities, just as we have related SZ intensity to the 
frequency-independent Compton $y$-parameter.
We define the frequency-independent cluster visibilities $Y(u,v)$ as
\begin{equation}
\label{eq:Yuv}
V_\nu(u,v) \equiv g(x) \, I_0 \, Y(u,v),
\end{equation}
where $I_0$ (in units of flux per solid angle) is
\begin{equation}
\label{eq:I0_cmb}
I_0 = \frac{2 (k_B \Tcmb)^3}{(h c)^2}.
\end{equation}
Additionally, we rescale $Y(u,v)$ by the square of the angular diameter distance, $d_A^2$, in order
to remove the redshift dependence from the cluster SZ signal.  
Note that, while we use the non-relativistic $g(x)$ (Eq.~\ref{eq:g_x}) to compute $Y(u,v)$ 
(Eq.~\ref{eq:Yuv}) , we only use this for display purposes.  The effects of assuming the classical
SZ frequency dependence are discussed in \S~\ref{profiles}.

\begin{figure}
\centerline{
\includegraphics[height=1.8in]{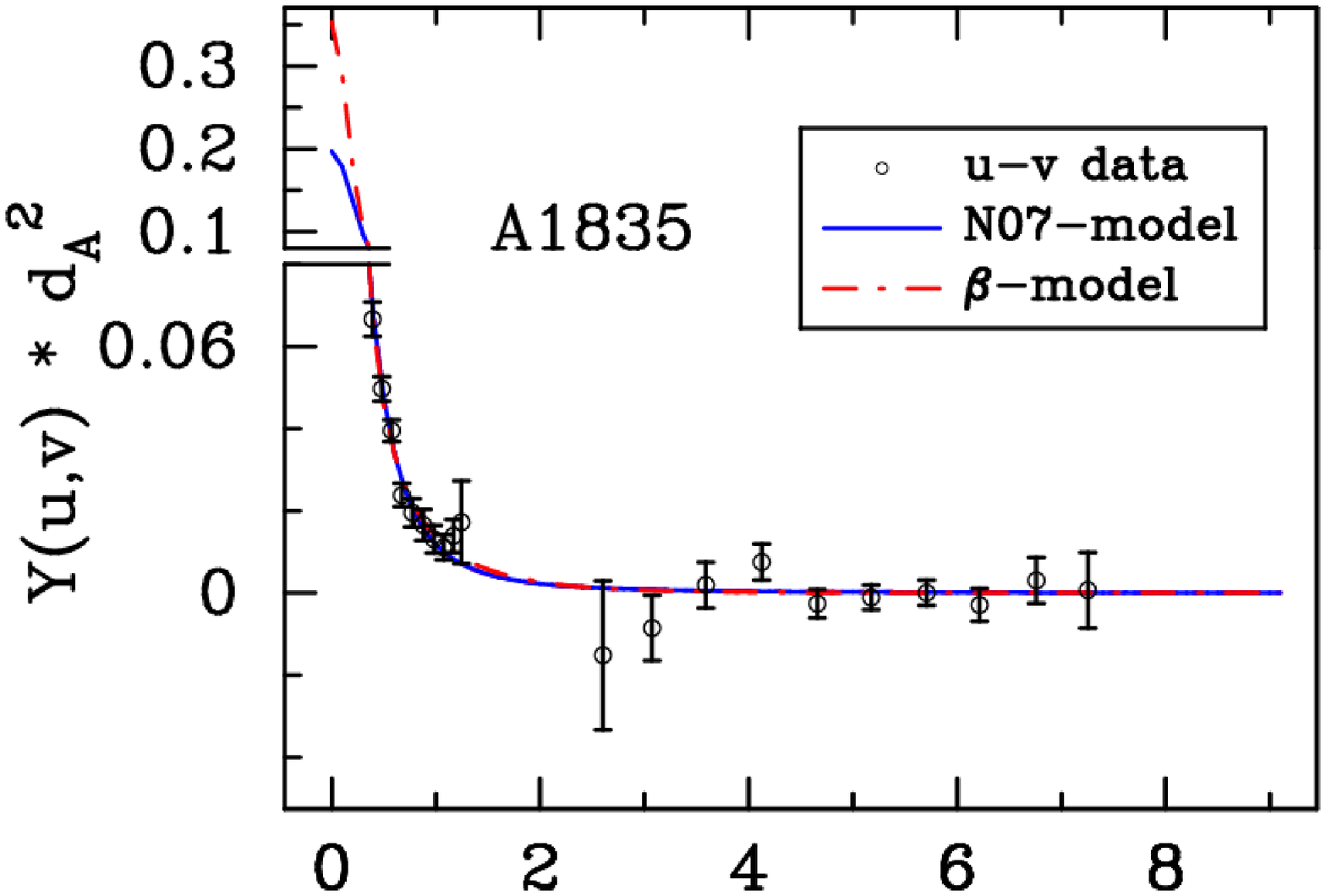}
\includegraphics[height=1.8in]{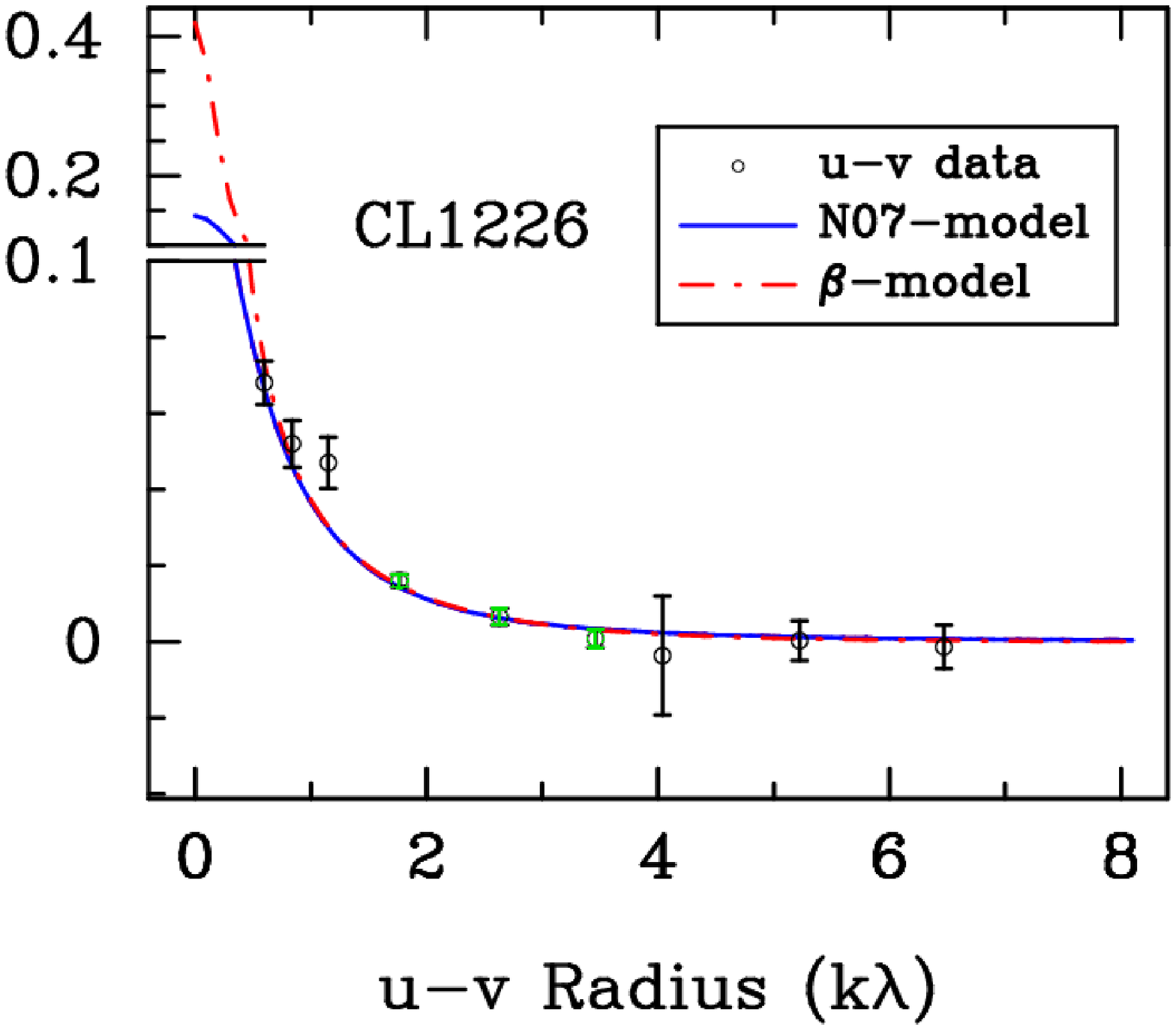}
\includegraphics[height=1.8in]{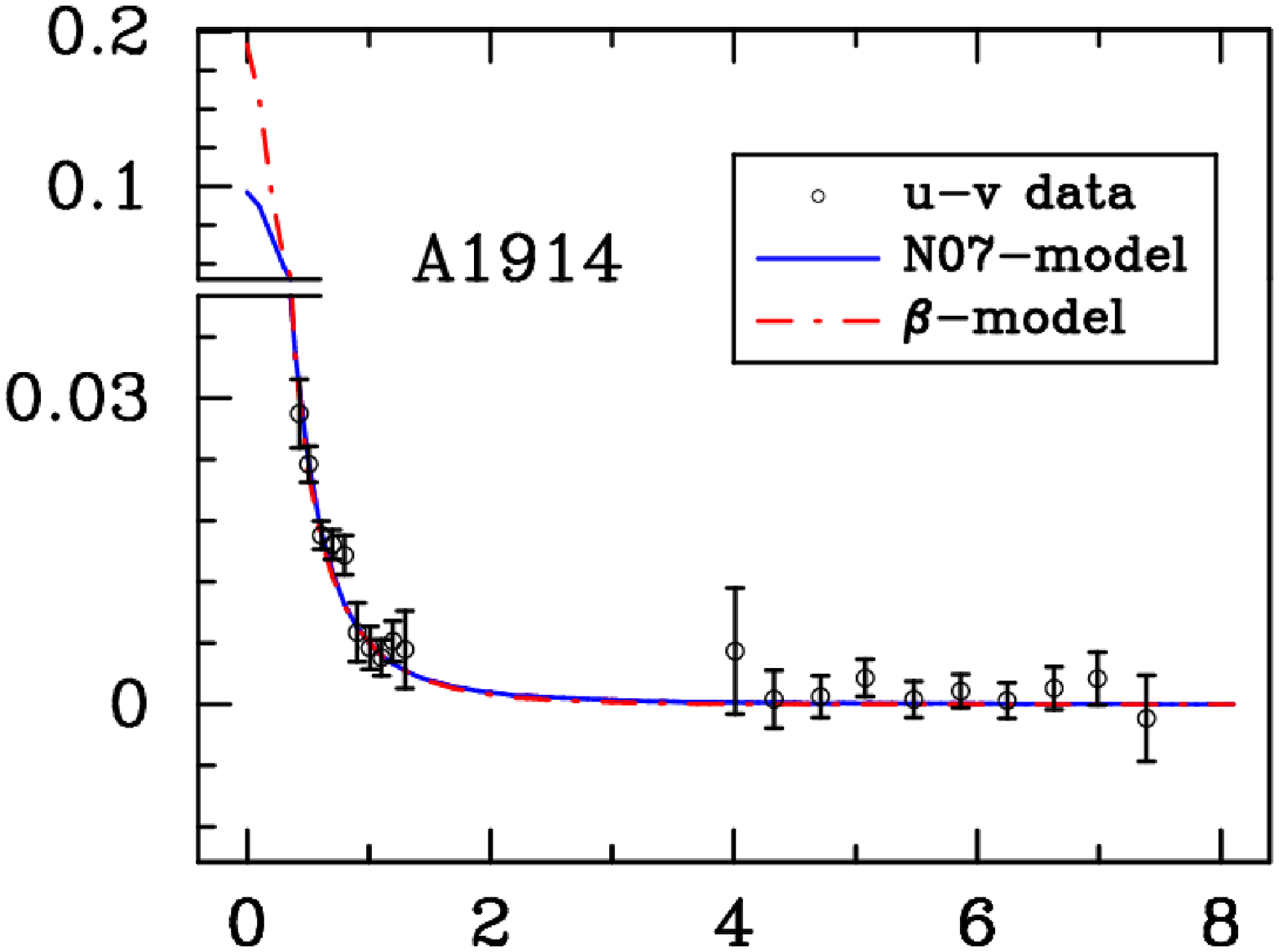}
}
\caption{SZ profiles for A1835 (left), CL1226 (middle), and A1914 (right),
plotted as a function of \emph{u,v}-radius ($\sqrt{u^2 + v^2}$).  
The panels show the real component of the visibilities over the full range of the fit models, 
using a broken axis to capture the model predictions as the \emph{u,v}-radius 
approaches zero k$\lambda$.
Using Eq.~\ref{eq:Yuv}, the visibilities were rescaled to $Y(u,v) \, d_A^2$, which removes
both the frequency and redshift dependence of the cluster visibilities.
The black points with error bars (1-$\sigma$) are the binned 30-GHz SZA visibility data.  
Models for the compact sources have been subtracted from the visibility
data before rescaling.  
The blue solid line is a high likelihood N07 model fit, while the red dashed line is a similarly-chosen
fit of the isothermal $\beta$-model.
For the available data points, both cluster models fit equally well.  However, note that as the 
\emph{u,v}-radius approaches zero k$\lambda$ (corresponding to large spatial scales on the sky)---where
 there are no data to constrain the models---the $\beta$-model predicts much more flux than the N07 model.
The middle three (green) data points for CL1226 are taken from 90-GHz
SZA observations. 
}\label{fig:uvplots}
\end{figure}

Figure~\ref{fig:uvplots} shows the maximum-likelihood fits of the 
N07 profile and the isothermal $\beta$-model to each cluster's visibility data,
from which we have subtracted the detected radio sources (Table~\ref{ptsrcTable}).  
We also removed the frequency dependence of the SZ effect by rescaling the 
cluster visibilities to $Y(u,v)$ (Eq.~\ref{eq:Yuv}).  
For the purposes of plotting, this rescaling is useful when binning the SZ signal 
across 8~GHz of bandwidth as well as when plotting the 30-GHz and 90-GHz SZA data 
taken on CL1226.

As indicated in Fig.~\ref{fig:uvplots}, both the isothermal $\beta$-model and N07 model (which was fit jointly 
with the SVM) fit the available data equally well.  However, as $\sqrt{u^2 + v^2} \rightarrow 0$ (large
scales on the sky), the isothermal $\beta$-model extrapolates to a much larger value of $Y(u,v)$.  
This corresponds to the much larger values of \Ycyl\ that are computed at large radii 
using fit $\beta$-models (see \S \ref{global}).

The middle panel of Fig.~\ref{fig:uvplots} shows the combined 30+90~GHz observations of CL1226,
which has a smaller angular extent than A1835 or A1914 due to its distance (compare, for example, the values
of $r_{500}$ for each cluster, listed in Table~\ref{table:derivedQuants_r500}).
The dynamic range and \emph{u,v}-space coverage provided by the 30-GHz SZA observations (black points) on CL1226 
were insufficient for constraining the radial pressure profile of the cluster.
The short baselines of the 90-GHz SZA observations (middle three points) help to provide
more complete \emph{u,v}-space coverage, as discussed in \S \ref{sza}.

\subsection{X-ray Surface Brightness Fits}\label{xray_surffits}

\begin{figure}
\begin{center}
\includegraphics[width=1.8in,angle=270]{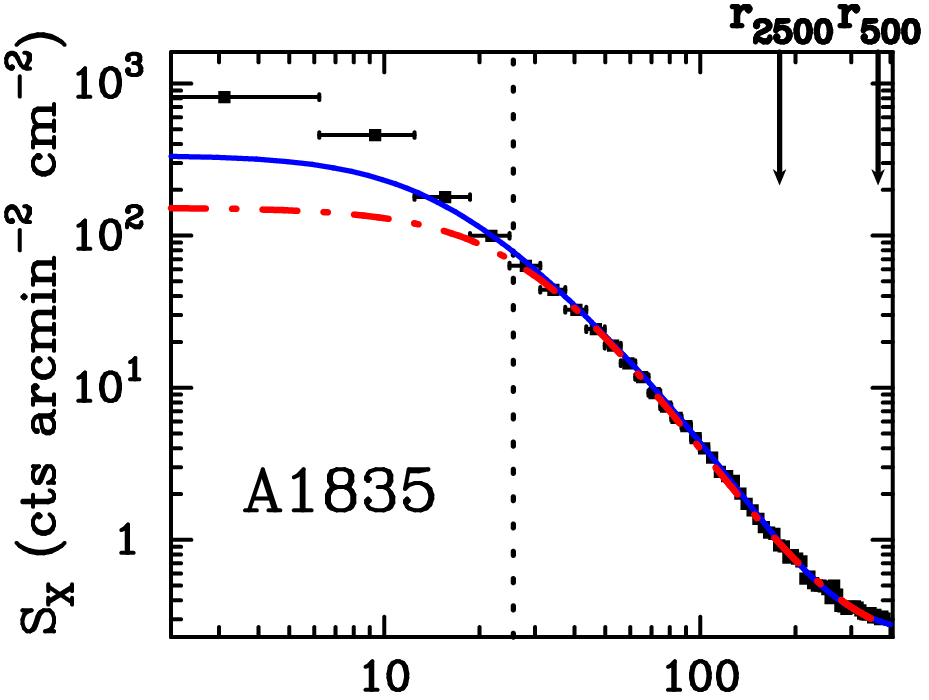}
\includegraphics[width=1.8in,angle=270]{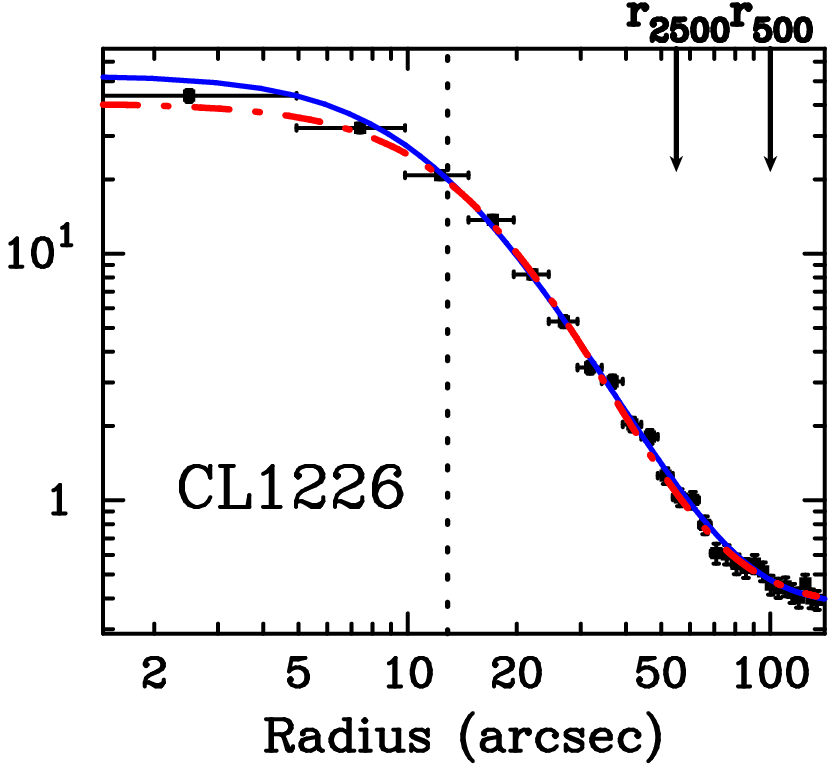}
\includegraphics[width=1.8in,angle=270]{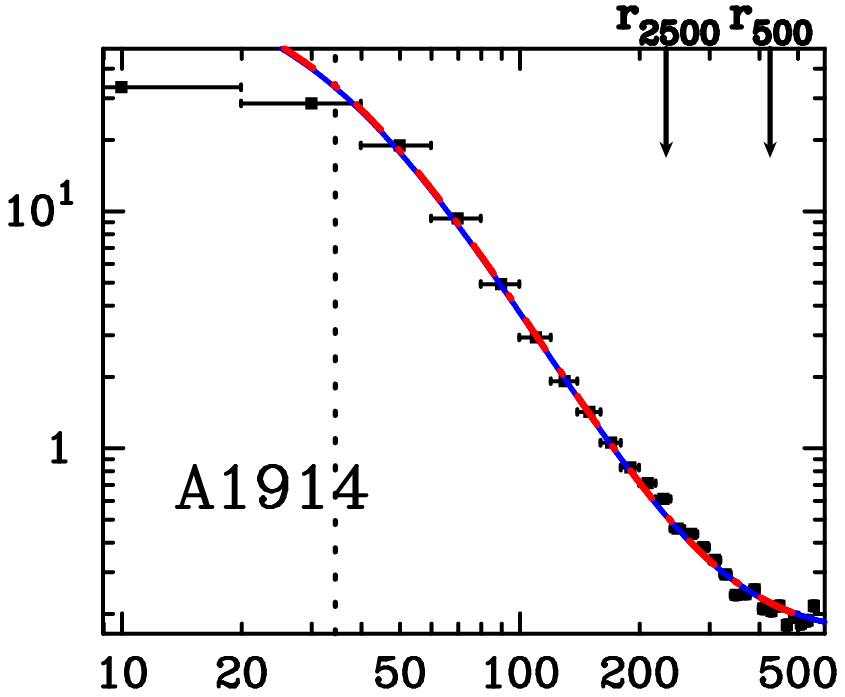}
\end{center}
\caption{X-ray surface brightness profiles for A1835 (left), CL1226 (middle), and A1914 (right).
The vertical dashed line denotes the 100~kpc core cut.
In each panel, the blue, solid line is the surface brightness computed using a high-likelihood fit of the 
N07+SVM profiles (analogous to the SZ models plotted in Fig.~\ref{fig:uvplots}), while the red, 
dot-dashed line is that from an isothermal $\beta$-model fit.  
Both fit lines include the X-ray background that was simultaneously fit with the cluster emission model.
The black squares are the annularly-binned X-ray data, where the widths of the bins are denoted
by horizonal error bars.  The vertical error bars are the 1-$\sigma$ errors on the measurements.
Arrows indicate $r_{2500}$ and $r_{500}$ derived from the N07+SVM profiles (see Tables 
\ref{table:derivedQuants_r2500} \& \ref{table:derivedQuants_r500}).}\label{fig:Sx_profiles}
\end{figure}

The X-ray surface brightness (Eq.~\ref{eq:xray_sb}), ignoring the data within a 100 kpc radius, 
was modeled separately with both the isothermal 
$\beta$-model, using the spectroscopically-determined, global $T_X$ (measured within 
$r \in [0.15,1.0] \, r_{500}$), and the SVM, using the temperature derived from the N07 pressure 
profile fit to the SZ data.  
Figure~\ref{fig:Sx_profiles} shows the maximum-likelihood fits to the surface
brightness of each cluster for both the SVM and isothermal $\beta$-model.  For plotting purposes, 
the X-ray data  are radially-averaged around the cluster centroid, which is determined in the 
joint SZ+X-ray analysis by fitting the two-dimensional X-ray imaging data with the 
spherically-symmetric SVM and isothermal $\beta$-model profiles.  

\subsection{ICM Profiles \label{profiles}}

\begin{figure}
\includegraphics[width=6in]{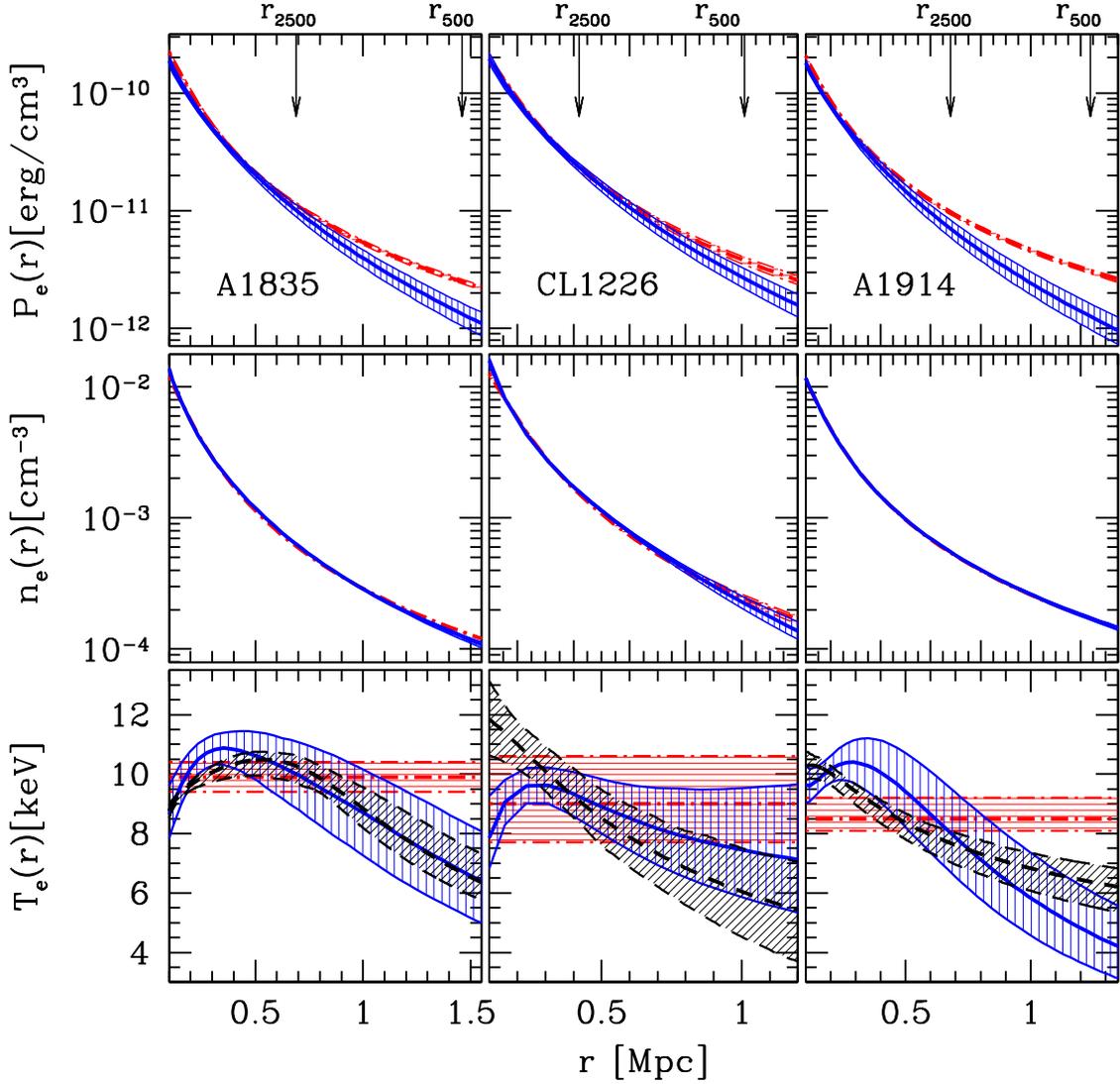}
\caption{Three-dimensional ICM radial profiles derived from the joint
analysis of SZA+{\em Chandra} X-ray surface brightness data for
A1835, CL1226, and A1914 (from left to right).  From top to
bottom: the electron pressure $P_e(r)$, the electron density $n_e(r)$,
the derived electron temperature $T_e(r)$, profiles as a function of the
cluster-centric radius.  The lines show the median deprojected
quantity derived from data using the N07+SVM (solid lines, vertically-hatched regions) and the
isothermal $\beta$-model (dot-dashed lines, horizontally-hatched regions).  The derived electron
temperature profiles are compared to the spectroscopically-determined
radial temperature profiles (black dashed lines, slanted hatching) obtained 
according to methods presented in M07/M08.
The hatched region indicate the 68\% confidence on parameters derived from each
model.  The arrows denote the median values of $r_{2500}$ and $r_{500}$ from the 
N07+SVM fits.}\label{fig:profiles}
\end{figure}

Figure~\ref{fig:profiles} shows the three-dimensional ICM radial
profiles derived from the joint analysis of SZA + \emph{Chandra} X-ray
surface brightness data for A1835, CL1226 and A1914 (from left to
right).  From top to bottom, we show the electron pressure, the gas
density and the derived electron temperature profiles, each as a function of cluster
radius.  In all panels, we compare the ICM profiles derived from the
N07+SVM model to the results of a traditional isothermal $\beta$-model
analysis, indicated by solid and dot-dashed lines, respectively.  The
hatched regions indicate the 68\% confidence interval for each derived
parameter.

As shown in the top panels of Fig.~\ref{fig:profiles}, the pressure
profiles derived from the N07 model and the isothermal $\beta$-model
show good agreement within $r_{2500}$, but deviate by
$\sim$3--5-$\sigma$ in the cluster outskirts.  This is a consequence
of the fact that clusters exhibit a significant decline in
temperature beyond $r_{2500}$, as determined from spectroscopic X-ray
observations (see the bottom panel of Fig.~\ref{fig:profiles}).  The
pressure profile derived from the isothermal $\beta$-model analysis
is therefore biased systematically high beyond $r_{2500}$.  In 
contrast, the N07 model, which fits the pressure directly, is free to
capture the true shape of the pressure profile well beyond the radius
at which the assumption of isothermality becomes invalid.

For all three clusters there is little evidence for the second
component of the electron density allowed by the SVM; the density is
fit equally well by either the SVM or a single-component
$\beta$-model, as illustrated by the center row of panels in
Fig.~\ref{fig:profiles} (see also Table~\ref{xrayTable}).  For all
three clusters, the fits of the SVM agree to within 1--2\% of the full V06
density profile fits (not shown) outside the core; discrepancies at
this level are easily attributed to differences in the fitting of the X-ray background, and
to the differences between the APEC and Raymond-Smith emissivity
models used respectively in the M07/M08 X-ray analysis and the joint SZ+X-ray
analyses.

In the bottom panels, the electron temperature profiles inferred from
the N07+SVM profiles are compared to temperature profiles derived from
deep spectroscopic \emph{Chandra} X-ray observations (and
\emph{XMM-Newton} in the case of CL1226; see M07).  The comparison
shows that the radial electron temperature profiles derived from the N07+SVM
profiles are in good agreement with independent X-ray measurements for
the relaxed clusters A1835 and CL1226, which exhibit radially
decreasing temperature profiles in the cluster outskirts \citep[see
also][]{markevitch1998,vikhlinin2005a}.  The disturbed cluster A1914,
however, shows less overall agreement between the derived N07+SVM radial
temperature profile and the M08 fit of the V06 temperature profile.
Since the N07 pressure profile fit to the SZA observation of A1914 agrees 
with that derived from M08 within their respective 68\% confidence intervals, 
the temperature discrepancy is likely due to deviations from the spherical symmetry
implicitly assumed in this analysis.
Additionally, scales greater than about six arcminutes are beyond the radial
extents probed by the SZA; it is unsurprising the agreement becomes poorer
at radii larger than this.

We note that we fit the SZ data using relativistic corrections to the SZ frequency dependence 
provided by \citet{itoh1998}.  These corrections are appropriate for the thermal SZ 
effect at the SZA observing frequencies of 30 and 90~GHz.
Compared to fits assuming the classical SZ frequency dependence, a pressure profile fit using the 
relativistic corrections has both a higher normalization and larger upper error bars.  
This is noticeable when including higher frequency data, where the relativistic correction is larger 
($\sim 5\%$ at 90~GHz versus $\sim 3\%$ at 30~GHz for cluster temperatures $\sim$~8~keV).
This increase in the pressure fit is due to the diminished magnitude of the SZ effect
when using these corrections (for frequencies below the null in the SZ spectrum, 
$\lesssim 218$~GHz; see e.g. \citet{itoh1998}). 
The pressure profile therefore must adjust to fit the observed SZ flux.

The larger upper error bars on the fit pressure profile arise from a more subtle effect.  Since the
temperature is derived from the simultaneously fit pressure and density profiles, and the 
SZ effect diminishes as electrons become more relativistic (i.e. hotter), the upper error
bar of the pressure fit must increase to fit the same noise in the observation (compared
to the non-relativistic case).
The lower error bar is less affected, as lower electron temperatures require smaller 
relativistic corrections.  The resulting asymmetric error bars can be seen in the derived 
temperature profile of CL1226, which relied on 90~GHz data, in Figure~\ref{fig:profiles}.


\subsection{Measurements of $Y$, \Mgas, \Mtot, and \fgas \label{global}}

In Tables \ref{table:derivedQuants_r2500} and
\ref{table:derivedQuants_r500}, we report global properties of
individual clusters derived from the N07+SVM model fits to the
SZ+X-ray data.  We calculate all quantities at overdensity radii
$r_{2500}$ and $r_{500}$, and compare them to results from both the
isothermal $\beta$-model analysis of the same data, as well as to the
X-ray-only analysis.

\begin{figure}
\centerline{\includegraphics[width=4in]{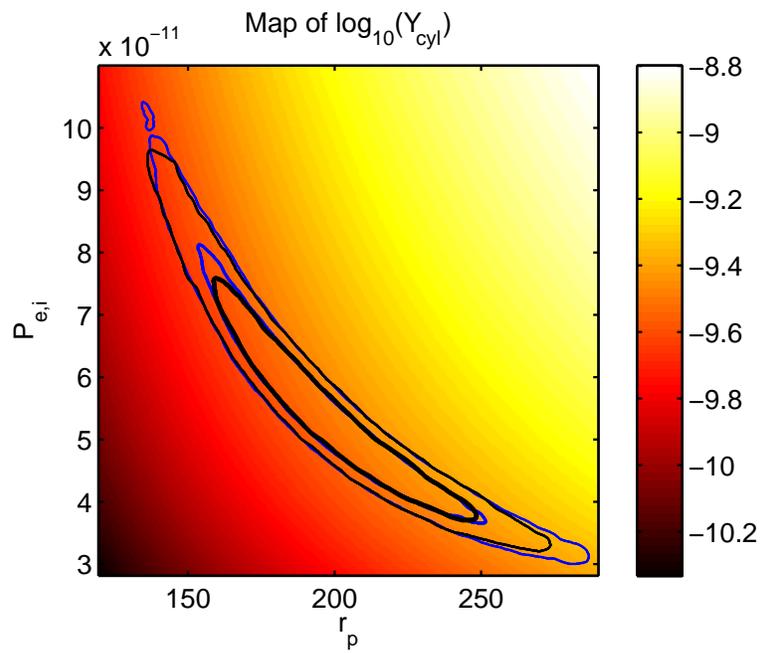}}
\caption{\Ycyl\ computed within 6\arcmin\ ($\sim 1.4~\rm Mpc$, which is $\approx r_{500}$
for this cluster) for fits to SZA observation of A1835. The bold, black contours contain 68\% and
95\% of the accepted iterations on the jointly-fit {\em Chandra} + SZA data, while the
thinner, blue contours are those for fits to SZA data alone.}\label{fig:a1835_yint_joint_SZ_degen}
\end{figure}

The N07 pressure profile has just two free parameters---$P_{e,i}$ and
$r_p$---which exhibit a degeneracy. Figure~\ref{fig:a1835_yint_joint_SZ_degen} shows this
degeneracy in fits of the N07 profile to the SZA observations of A1835. 
Similar to the $r_c-\beta$ degeneracy of the $\beta$-model
\citep[see][for example]{grego2001}, these two quantities are not
individually constrained by our SZ observations, but they are tightly
correlated and the preferred region in the $P_{e,i}-r_p$ plane encloses
approximately constant \Ycyl. As a result, the 68\% confidence region for \Ycyl\
is more tightly constrained than the large variation in $P_{e,i}$ or $r_p$ might 
individually indicate. 
Figure~\ref{fig:a1835_yint_joint_SZ_degen} also shows that the inclusion of
X-ray data has only a marginal effect on the value of \Ycyl\ derived from the SZ fit. This is
as expected, due to
the weak dependence of the X-ray surface brightness on temperature (see \S~\ref{xray_sze})
and the fact that the N07 profile is not linked to the SVM density profile's shape.
This indicates that X-ray data are not necessary to constrain \Ycyl, although they do limit
the range of accepted radial pressure profiles.

At both $r_{2500}$ and $r_{500}$, the measurements of \Ycyl\ derived from the joint
N07+SVM and the X-ray-only analysis are consistent at the 1-$\sigma$
level, for all three clusters.  The isothermal $\beta$-model analysis,
however, overestimates \Ycyl\ by $\sim$~20\%--40\% at $r_{2500}$, and
by $\sim$~30\%--115\% at $r_{500}$.  This is due to the large
contribution to \Ycyl\ from the cluster outskirts, at every projected radius, 
where the $\beta$-model significantly overestimates the
pressure.  

In contrast to the systematic variations in \Ycyl, the determinations of $\Ysph(r_{2500})$ are
consistent among the three analyses.  At $r_{500}$, however, the
median \Ysph\ values from the isothermal $\beta$-model can be as much as
$\sim$ 60\% higher than either the N07+SVM or M08 results, due to
the fact that isothermality is a poor description of the cluster
outskirts.

The gas mass estimates computed using either the jointly-fit SVM or
the isothermal $\beta$-model agree with the gas masses derived from
the Maughan X-ray fits (Tables \ref{table:derivedQuants_r2500} and
\ref{table:derivedQuants_r500}).  This agreement is not surprising,
given that the gas mass is determined in all cases from density fits
to the X-ray data.  It demonstrates, however, that the 100~kpc core
makes a negligible contribution to the total gas mass even at
$r_{2500}$, and that excluding the core from the joint analysis does
not therefore introduce any significant bias in our estimate of
\Mgas. Incidentally, it also shows that the additional component
allowed by the SVM and the full V06 density models is not indicated in
these clusters.

In Tables \ref{table:derivedQuants_r2500} and
\ref{table:derivedQuants_r500}, we also present estimates of the total
masses, computed using each model's estimate of the overdensity radius
($r_\Delta$) for each Monte Carlo realization of the fit parameters.
For two of the clusters, we find that the error bars are significantly
larger for \Mtot\ determined from the N07+SVM fits than for the
isothermal $\beta$-model or M08 fits.  This is a consequence of the fact
that the $\beta$-model analysis, with fewer free parameters and the
assumption of isothermality, effectively places stronger but poorly-justified
priors on the total mass.
We find that the N07+SVM and M08 total mass estimates agree
at both $r_{500}$ and $r_{2500}$, leading to good overall agreement
between gas fractions computed using the N07+SVM profiles and those
from the Maughan X-ray fits.

\begin{figure}
\centerline{\includegraphics{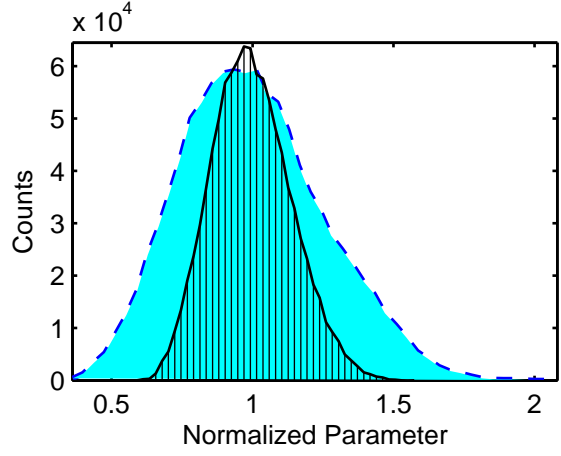}}
\caption{1-D histograms of \Mtot\ (cyan/light gray region
with dashed outline) and \Ycyl\ (vertically-hatched region with solid outline), normalized by their
respective median values, derived from the joint fits of the N07+SVM profiles to A1835. 
Both \Mtot\ and \Ycyl\ are computed within a fixed radius of $\theta=6\arcmin$.  
The derived \Ycyl, which scales directly with integrated SZ flux, has a more tightly constrained 
and centrally peaked distribution than that of \Mtot, as \Mtot\ is sensitive to the change 
in slope in the pressure profile ($dP/dr$).}\label{fig:1dhists}
\end{figure}

Since the HSE estimate for \Mtot\ is sensitive to the change in slope in the pressure 
profile ($dP/dr$), \Mtot\ is not as well constrained by the N07+SVM profiles as is \Ycyl,
which scales directly with integrated SZ flux, a parameter that is more directly linked 
to what the SZA observes (see \S~\ref{sze_visfits}).
Figure~\ref{fig:1dhists} shows a comparison of the N07+SVM estimates for \Mtot\ and \Ycyl, 
revealing that \Ycyl\ has a more tightly constrained and centrally peaked distribution than 
\Mtot\ does. 

\begin{figure}
\includegraphics[width=6in]{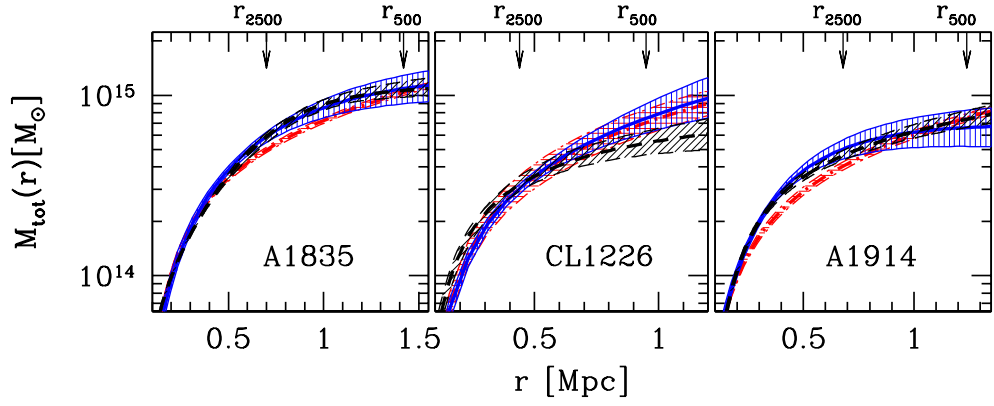}
\caption{Total mass profiles derived from the joint analysis of
SZA+{\em Chandra} X-ray surface brightness data for A1835,
CL1226, and A1914 (from left to right).  The line types are the same
as those in Fig.~\ref{fig:profiles}.}\label{fig:massprofiles}
\end{figure}

The isothermal $\beta$-model, on the other hand, is over-constrained such that
it cannot agree with the non-isothermal estimates of \Mtot\ at both $r_{2500}$ and $r_{500}$,
Its estimates of \Mtot\ are moreover sensitive to the annulus within
which $T_X$ is determined (see \S~\ref{sec:xray}).  This trend can be
seen in Fig.~\ref{fig:massprofiles}, which shows \Mtot$(r)$ for each
cluster.


\section{Conclusions}\label{conclusions}
In this work, we demonstrated the application of a new pressure
profile, proposed by \citet{nagai2007b}, in fitting SZ data taken by the Sunyaev-Zel'dovich Array.  
By combining the pressure profile constraints from the SZ data with constraints on density from \emph{Chandra} X-ray 
imaging, using a simplified form of the density model proposed by \citet{vikhlinin2006}, 
we were able to determine the properties of three clusters (A1835,  A1914, and CL~J1226.9+3332) 
spanning a wide range in redshift and dynamical states. 

This technique was compared with two others:
a joint analysis of the same SZ + X-ray data relying on the commonly used isothermal $\beta$-model,
which assumes the pressure and density profiles have the same form,
and an X-ray only analysis that independently models temperature and density 
using both X-ray imaging and spectroscopic measurements (but excluding SZ data).
We find that the global cluster properties at both $r_{2500}$
and $r_{500}$ determined from ICM profiles fit in the joint pressure analysis are in excellent agreement
with properties determined in the X-ray analysis.
By contrast, the isothermal $\beta$-model tends to
overestimate with respect to the other models the gas pressure at $r > r_{2500}$, where
isothermality is an increasingly poor assumption. The $\beta$-model
thus leads to an overestimate of the cylindrically-integrated Compton $y$-parameter
at $r_{500}$.
Since the isothermal $\beta$-model does not provide a good description of
the ICM profile in the cluster outskirts, we caution
against its use in deriving global properties of clusters even at $r \sim r_{500}$,
which is a large fraction of the virial radius.

We tested the ability to recover the ICM electron pressure profiles 
from SZ data by analyzing the SZ data together with the X-ray imaging data alone, 
ignoring the X-ray spectroscopic information.
Assuming the ideal gas law, we derive electron temperature profiles by
coupling the pressure fits to the SZ data with the density fits to the X-ray imaging data.
We find that these derived temperature profiles show broad agreement with those
determined spectroscopically from deep X-ray observations, even for
the highest redshift cluster in our sample, at $z=0.89$.  This method
therefore provides an independent technique for determining the radial
electron temperature distribution in high-redshift clusters, for which
deep spectroscopic X-ray data may be unavailable and are impractical to
obtain.


\acknowledgments

We thank John Cartwright, Ben Reddall and Marcus Runyan for their  significant contributions 
to the construction and commissioning of the \facility{SZA} instrument.  We thank Esra Bulbul and 
Nicole Hasler for their insights, comments, and help with \emph{Chandra} X-ray Observatory 
(\facility{CXO}) data reduction and figure production. 
We thank the staff of the Owens Valley Radio Observatory and CARMA for their outstanding support.
We thank Samuel LaRoque for his help with the modeling code.
We gratefully acknowledge the James S.\ McDonnell Foundation, the National Science Foundation 
and the University of Chicago for funding to construct the SZA.  
The operation of the SZA is supported by NSF Division of Astronomical Sciences through
grant AST-0604982. Partial support is provided by NSF Physics Frontier Center grant PHY-0114422 
to the Kavli Institute of Cosmological Physics at the University of Chicago, and by NSF grants 
AST-0507545 and AST-05-07161 to Columbia University.  AM acknowledges support from a Sloan Fellowship, 
DM from an NRAO Jansky Fellowship, BM from a Chandra Fellowship, and CG, SM, and MS from NSF Graduate Research Fellowships.

\bibliography{genNFW}
\begin{deluxetable}{lcc|cc|cc|cc|c}
\tabletypesize{\scriptsize}
\tablecolumns{8}
\setlength{\tabcolsep}{2.15mm}
\tablecaption{SZA Cluster Observations}
\tablehead{
\colhead{Cluster Name}& \colhead{$z$\tablenotemark{a}}& \colhead{$D_A$} &  
\multicolumn{2}{c}{\underline{SZA Pointing Center (J2000)}}&
\multicolumn{2}{c}{\underline{Short Baselines (0.3--1.5~$\rm k\lambda$)}}&
\multicolumn{2}{c}{\underline{Long Baselines (3.0--7.5~$\rm k\lambda$)}} &
\colhead{$\rm{\tau_{int,red}}$\tablenotemark{b}}
\\
\colhead{} & \colhead{} & \colhead{(Gpc)} & 
\colhead{$\alpha$} & \colhead{$\delta$} & 
\colhead{beam($\arcsec\times\arcsec$)\tablenotemark{c}} & \colhead{$\sigma$(mJy)\tablenotemark{d}} &
\colhead{beam($\arcsec\times\arcsec$)\tablenotemark{c}} & \colhead{$\sigma$(mJy)\tablenotemark{d}} &
\colhead{(hrs)}
}
\startdata
A1914          & 0.17 & 0.60 & $14^h 26^m 00^s\!.8$ & $+37^{\circ}49^{\prime}35\arcsec\!.7$ & 117.5$\times$129.9 & 0.30 & 23.5$\times$17.4 & 0.35 & 11.5\\
A1835          & 0.25 & 0.81 & $14^h 01^m 02^s\!.0$ & $+02^{\circ}52^{\prime}41\arcsec\!.7$ & 116.6$\times$152.1 & 0.25 & 17.5$\times$23.5 & 0.33 & 18.6\\
CL1226 (30-GHz)& 0.89 & 1.60 & $12^h 26^m 58^s\!.0$ & $+33^{\circ}32^{\prime}45\arcsec\!.0$ & 117.4$\times$125.4 & 0.20 & 16.0$\times$21.2 & 0.20 & 22.0\\
CL1226 (90-GHz)\tablenotemark{e}& " & " & " & "                                             & 42.3$\times$39.1 & 0.42 & -- & -- &29.2\\
\enddata

\label{obsTable}

\tablenotetext{a}{Redshifts for A1914 and A1835 are from \cite{struble1999}. 
Redshift for CL1226 is from \cite{ebeling2001}. 
All are in agreement with XSPEC fits to iron emission lines presented in \cite{laroque2006}.}
\tablenotetext{b}{Unflagged data after reduction.}
\tablenotetext{c}{Synthesized beam FWHM and position angle measured from North through East}
\tablenotetext{d}{Achieved {\em rms} noise in corresponding maps}
\tablenotetext{e}{The short baselines of the 90-GHz observation probe $\sim$1--4.5 $\rm k\lambda$.  The long
baselines of the 90-GHz observation were not used here.}

\end{deluxetable}

\begin{deluxetable}{cc|cc|cc}
\tabletypesize{\footnotesize}
\tablecolumns{8}
\setlength{\tabcolsep}{1.4mm}
\tablecaption{Details of X-ray Observations}
\tablehead{
\colhead{Cluster Name} &  \colhead{ObsID} & 
\multicolumn{2}{c}{\underline{Joint SZ+X-ray Analysis}}&
\multicolumn{2}{c}{\underline{Maughan X-ray Analysis}}
\\
 &  & 
\colhead{$\rm{\tau_{int}}$\tablenotemark{a}} &\colhead{fitting region\tablenotemark{b}} &
\colhead{$\rm{\tau_{int}}$\tablenotemark{a}} &\colhead{fitting region}
\\
 &  &
\colhead{(ks)} & \colhead{\arcsec} &
\colhead{(ks)} & \colhead{\arcsec} 
}
\startdata
A1914	& 542+3593	& 26.0  & 34.4--423.1 & 23.3 & 0.0--462.5 \\
A1835	& 6880		& 85.7  & 25.6--344.4 & 85.7 & 0.0--519.6 \\
CL1226	& 3180+5014     & 64.4  & 12.9--125.0 & 50   & 0.0--125.0 \\
" 	& 0200340101 	& ~~N/A & N/A         & 75+68\tablenotemark{c} & 17.1--115\tablenotemark{c}
\enddata

\label{xrayTable}

\tablenotetext{a}{Good (unflagged, cleaned) times for X-ray observations after respective calibration pipelines.}
\tablenotetext{b}{X-ray image fitting region.  Excluded regions are not used in the X-ray likelihood calculation (Eq.~\ref{eq:L_image}).}
\tablenotetext{c}{{\em XMM-Newton} observation of CL1226, presented in M07, was used only in the spectroscopic analysis.
The exposure times are, respectively, those of the MOS and PN camera.}
\end{deluxetable}

\begin{deluxetable}{lc|ccc}
\tabletypesize{\footnotesize}
\tablecolumns{6}
\setlength{\tabcolsep}{2.15mm}
\tablecaption{Unresolved Radio Sources in 30-GHz Observations}
\tablehead{
\colhead{\underline{Cluster Name}}& 
\colhead{\underline{Src \#}}&  
\multicolumn{2}{c}{\underline{Differential Offset from Pointing Center}}&
\colhead{\underline{Flux at 30.938~GHz}}
\\
\colhead{} & \colhead{} & 
\colhead{$\Delta$ R.A. (\arcsec)} & \colhead{$\Delta$ Dec. (\arcsec)} & 
\colhead{mJy}   
}
\startdata
A1914 	& 1 & -242.1 & -235.4 & $1.8^{+0.4}_{-0.4}$  \\
A1914 	& 2 & -160.5 & -108.6 & $1.2^{+0.3}_{-0.3}$  \\
A1835	& 1 & ~~~0.9 & ~~~1.5 & 2.8$^{+0.3}_{-0.3}$  \\
A1835	& 2 & ~-22.8 & ~-51.2 & 1.1$^{+0.4}_{-0.3}$  \\
A1835	& 3 & -275.1 & -178.6 & 0.8$^{+0.4}_{-0.4}$  \\
CL1226\tablenotemark{a}	& 1 & ~258.4 & ~-38.9 & $3.9^{+0.2}_{-0.2}$ \\
\enddata

\label{ptsrcTable}

\tablenotetext{a}{No point sources were detected in the 90-GHz observation of CL1226.}
\end{deluxetable}

\begin{deluxetable}{l|cc|cc|ccc}
\renewcommand{\arraystretch}{0.85}
\tabletypesize{\footnotesize}
\tablecolumns{8}
\setlength{\tabcolsep}{2.1mm}
\tablecaption{$Y_{\rm cyl}$, $Y_{\rm sph}$, $M_{\rm gas}$, $M_{\rm tot}$, and $f_{\rm gas}$ for Each Model Computed Within $r_{2500}$.}
\tablehead{
Cluster   
& {$\theta_{2500}$} &  {$r_{2500}$} & {$Y_{\rm cyl}$} & {$Y_{\rm sph}$} &{$M_{\rm gas}$} &{$M_{\rm tot}$} &{$f_{\rm gas}$} \\
model fit 
& (\arcsec) & (Mpc) & {($10^{-5} {\rm Mpc}^2$)}  & {($10^{-5} {\rm Mpc}^2$)} &{($10^{13} {\rm M_\odot}$)} &{($10^{14} {\rm M_\odot}$)} 
} 
\startdata

\underline{\bf Abell 1835} & & & & & & & \\[.25pc]
~N07+SVM              & 173$^{+5.8}_{-5.7}$   & 0.68$^{+0.02}_{-0.02}$ & 11.73$^{+1.36}_{-1.27}$ & 8.25$^{+0.81}_{-0.78}$ & 5.74$^{+0.26}_{-0.25}$ & 5.64$^{+0.58}_{-0.54}$ & 0.102$^{+0.006}_{-0.005}$ \\[.25pc]
~Maughan (this work)  & 169$^{+5.5}_{-8.0}$   & 0.66$^{+0.02}_{-0.03}$ & 11.58$^{+0.61}_{-0.67}$ & 7.88$^{+0.49}_{-0.72}$ & 5.77$^{+0.25}_{-0.35}$ & 5.30$^{+0.53}_{-0.72}$ & 0.109$^{+0.009}_{-0.006}$ \\[.25pc]
~isothermal $\beta$-model    & 159$^{+3.0}_{-2.9}$   & 0.62$^{+0.01}_{-0.01}$ & 13.85$^{+0.72}_{-0.67}$ & 7.94$^{+0.43}_{-0.40}$ & 4.96$^{+0.13}_{-0.12}$ & 4.38$^{+0.25}_{-0.24}$ & 0.113$^{+0.004}_{-0.004}$ \\
 & & & & & & & \\
\underline{\bf CL~J1226+3332.9} & & & & & & & \\[.25pc]
~N07+SVM              & 52.8$^{+1.8}_{-1.9}$  & 0.41$^{+0.01}_{-0.01}$ & ~5.49$^{+0.53}_{-0.50}$ & 3.56$^{+0.36}_{-0.36}$ & 2.91$^{+0.14}_{-0.14}$ & 2.67$^{+0.29}_{-0.27}$ & 0.109$^{+0.007}_{-0.007}$ \\[.25pc]
~\citet{maughan2007b} & 57.3$^{+1.6}_{-1.5}$  & 0.45$^{+0.01}_{-0.01}$ & ~7.57$^{+0.33}_{-0.34}$ & 5.04$^{+0.31}_{-0.28}$ & 3.25$^{+0.14}_{-0.13}$ & 3.41$^{+0.30}_{-0.26}$ & 0.095$^{+0.004}_{-0.004}$ \\[.25pc]
~isothermal $\beta$-model    & 54.2$^{+5.1}_{-4.3}$  & 0.42$^{+0.04}_{-0.03}$ & ~6.76$^{+0.91}_{-0.73}$ & 3.92$^{+0.60}_{-0.48}$ & 2.93$^{+0.40}_{-0.33}$ & 2.89$^{+0.90}_{-0.63}$ & 0.101$^{+0.014}_{-0.014}$ \\
 & & & & & & & \\
\underline{\bf Abell 1914} & & & & & & & \\[.25pc]
~N07+SVM              & 228$^{+12.9}_{-11.7}$ & 0.67$^{+0.04}_{-0.03}$ & ~8.43$^{+1.55}_{-1.24}$ & 6.29$^{+1.03}_{-0.82}$ & 4.71$^{+0.34}_{-0.30}$ & 4.97$^{+0.89}_{-0.72}$ & 0.095$^{+0.009}_{-0.009}$ \\[.25pc]
~\citet{maughan2008}  & 218$^{+7.1~}_{-5.7~}$ & 0.63$^{+0.02}_{-0.02}$ & ~7.87$^{+0.56}_{-0.55}$ & 5.69$^{+0.37}_{-0.38}$ & 4.64$^{+0.17}_{-0.16}$ & 4.31$^{+0.43}_{-0.33}$ & 0.107$^{+0.005}_{-0.006}$ \\[.25pc]
~isothermal $\beta$-model    & 204$^{+5.7~}_{-5.1~}$ & 0.59$^{+0.02}_{-0.01}$ & 11.28$^{+0.59}_{-0.56}$ & 6.24$^{+0.34}_{-0.32}$ & 4.04$^{+0.15}_{-0.14}$ & 3.52$^{+0.30}_{-0.26}$ & 0.115$^{+0.005}_{-0.005}$ \\
\enddata
\label{table:derivedQuants_r2500}
\end{deluxetable}

\begin{deluxetable}{l|cc|cc|ccc}
\renewcommand{\arraystretch}{0.85}
\tabletypesize{\footnotesize}
\tablecolumns{8}
\setlength{\tabcolsep}{2.1mm}
\tablecaption{$Y_{\rm cyl}$, $Y_{\rm sph}$, $M_{\rm gas}$, $M_{\rm tot}$, and $f_{\rm gas}$ for Each Model Computed Within $r_{500}$.}
\tablehead{
Cluster   
& {$\theta_{500}$} & {$r_{500}$}  & {$Y_{\rm cyl}$} & {$Y_{\rm sph}$} &{$M_{\rm gas}$}  &{$M_{\rm tot}$}  &{$f_{\rm gas}$} \\
model fit 
& (\arcsec) & (Mpc) &{($10^{-5} {\rm Mpc}^2$)}  & {($10^{-5} {\rm Mpc}^2$)} &{($10^{13} {\rm M_\odot}$)} &{($10^{14} {\rm M_\odot}$)} 
} 
\startdata

\underline{\bf Abell 1835} & & & & & & & \\[.25pc]
~N07+SVM              & 369$^{+28}_{-27}$ & 1.44$^{+0.11}_{-0.10}$ & 20.79$^{+3.79}_{-3.34}$ & 17.55$^{+3.00}_{-2.70}$ & 13.67$^{+1.03}_{-1.01}$ & 11.00$^{+2.68}_{-2.22}$ & 0.124$^{+0.020}_{-0.017}$ \\[.25pc]
~Maughan (this work)  & 363$^{+17}_{-12}$ & 1.42$^{+0.07}_{-0.05}$ & 21.37$^{+2.45}_{-1.58}$ & 17.41$^{+1.61}_{-0.99}$ & 13.94$^{+0.64}_{-0.52}$ & 10.68$^{+1.54}_{-1.01}$ & 0.133$^{+0.009}_{-0.012}$\\[.25pc]
~isothermal $\beta$-model    & 361$^{+7~}_{-6~}$ & 1.41$^{+0.03}_{-0.03}$ & 34.53$^{+1.78}_{-1.68}$ & 21.29$^{+1.09}_{-1.02}$ & 13.29$^{+0.27}_{-0.27}$ & 10.30$^{+0.58}_{-0.54}$ & 0.129$^{+0.005}_{-0.005}$ \\
 & & & & & & & \\
\underline{\bf CL~J1226+3332.9} & & & & & & & \\[.25pc]
~N07+SVM              & 127$^{+13}_{-10}$ & 0.98$^{+0.10}_{-0.07}$ & 11.9$^{+2.0}_{-1.6}$ & ~9.71$^{+1.58}_{-1.29}$ & 8.50$^{+0.68}_{-0.60}$ & ~7.37$^{+2.50}_{-1.57}$ & 0.115$^{+0.022}_{-0.023}$\\[.25pc]
~\citet{maughan2007b} & 115$^{+3~}_{-3~}$ & 0.89$^{+0.02}_{-0.02}$ & 13.9$^{+1.3}_{-1.1}$ & 10.59$^{+0.69}_{-0.68}$ & 8.30$^{+0.32}_{-0.36}$ & ~5.49$^{+0.46}_{-0.47}$ & 0.151$^{+0.008}_{-0.008}$\\[.25pc]
~isothermal $\beta$-model    & 126$^{+11}_{-9}$ & 0.98$^{+0.09}_{-0.07}$ & 16.8$^{+1.9}_{-1.6}$ & 10.97$^{+1.30}_{-1.06}$ & 8.21$^{+0.77}_{-0.65}$ & ~7.30$^{+2.10}_{-1.51}$ & 0.113$^{+0.018}_{-0.018}$\\
 & & & & & & & \\
\underline{\bf Abell 1914} & & & & & & & \\[.25pc]
~N07+SVM                    & 430$^{+38}_{-33}$ & 1.25$^{+0.11}_{-0.10}$ & 12.90$^{+2.97}_{-2.31}$ & 11.05$^{+2.44}_{-1.91}$ & 10.26$^{+0.94}_{-0.85}$ & ~6.62$^{+1.90}_{-1.42}$ & 0.155$^{+0.026}_{-0.023}$\\[.25pc]
~\citet{maughan2008}        & 448$^{+24}_{-21}$ & 1.29$^{+0.07}_{-0.06}$ & 13.47$^{+1.68}_{-1.77}$ & 10.78$^{+1.03}_{-1.09}$ & 10.24$^{+0.45}_{-0.57}$ & ~7.49$^{+1.29}_{-1.00}$ & 0.138$^{+0.015}_{-0.018}$\\[.25pc]
~isothermal $\beta$-model    & 461$^{+13}_{-11}$ & 1.34$^{+0.04}_{-0.03}$ & 29.08$^{+1.52}_{-1.44}$ & 17.07$^{+0.87}_{-0.84}$ & 11.05$^{+0.36}_{-0.33}$ & ~8.14$^{+0.69}_{-0.59}$ & 0.136$^{+0.007}_{-0.007}$\\
\enddata
\label{table:derivedQuants_r500}
\end{deluxetable}

\pagebreak

\end{document}